\documentclass[amsmath,amssymb,amsfonts,superscriptaddress,floatfix,twocolumn]{revtex4-1}

\usepackage[english]{babel}
\usepackage[colorlinks=true,citecolor=green,linkcolor=red,urlcolor=blue]{hyperref}
\usepackage{braket}
\usepackage{bbm}
\usepackage{graphicx}
\usepackage{color}
\usepackage{appendix}                                             
\usepackage[utf8]{inputenc}

 \usepackage{bbold}
 \usepackage[normalem]{ulem}

\DeclareMathOperator{\Tr}{Tr}

\newcommand{\Br}[1]{\left[#1\right]}
\newcommand{\unity}{\ensuremath{\mathbbm{1}}}

\begin{document}
\title{Truncating an exact Matrix Product State for the XY model: \\ transfer matrix and its renormalisation}
\author{Marek M. Rams}
\affiliation{Institute of Physics, Jagiellonian University,  \L{}ojasiewicza 11, 30-348 Krak\'ow, Poland }
\affiliation{Institute of Physics, Krak\'ow University of Technology, Podchora\.zych 2, 30-084 Krak\'ow, Poland }

\author{Valentin Zauner}
\affiliation{Vienna Center for Quantum Science and Technology, Faculty of Physics, University of Vienna, Vienna, Austria}

\author{Matthias Bal}
\affiliation{Ghent University, Krijgslaan 281, 9000 Gent, Belgium}

\author{Jutho Haegeman}
\affiliation{Ghent University, Krijgslaan 281, 9000 Gent, Belgium}

\author{Frank Verstraete}
\affiliation{Vienna Center for Quantum Science and Technology, Faculty of Physics, University of Vienna, Vienna, Austria}
\affiliation{Ghent University, Krijgslaan 281, 9000 Gent, Belgium}

\begin{abstract}
We discuss how to analytically obtain an -- essentially infinite -- Matrix Product State (MPS) representation of the ground state of the XY model. 
On the one hand this allows to illustrate how the Ornstein-Zernike form of the correlation function emerges in the exact case using standard MPS language. 
On the other hand we study the consequences of truncating the bond dimension of the exact MPS, which is also part of many tensor network algorithms, and analyze how the truncated MPS transfer matrix is representing the dominant part of the exact quantum transfer matrix. 
In the gapped phase we observe that the correlation length obtained from a truncated MPS approaches the exact value following a power law in effective bond dimension. In the gapless phase we find a good match between a state obtained numerically from standard MPS techniques with finite bond dimension, and a state obtained by effective finite imaginary time evolution in our framework. This provides a direct hint for a geometric interpretation of Finite Entanglement Scaling at the critical point in this case. 
Finally, by analyzing the spectra of transfer matrices, we support the interpretation put forward by [V. Zauner {\it at. al}., New J. Phys. {\bf 17}, 053002 (2015)] that the MPS transfer matrix emerges from the quantum transfer matrix though the application of Wilson's Numerical Renormalisation Group along the imaginary-time direction.
\end{abstract}

\maketitle


\section{Introduction}

Over the recent decades, Matrix Product States (MPS) \cite{Fannes1992,Verstraete2008,Schollwock2011} and related numerical techniques have become the standard framework for simulating low energy states of local Hamiltonians in 1D. 
MPS with finite bond dimension provide an exact representation of the ground state only for a certain well-designed class of so-called parent Hamiltonians (such as the celebrated AKLT model \cite{AKLT1987}). Nevertheless, for generic local gapped Hamiltonians, MPS of finite bond dimension approximate local quantities in the ground state essentially to arbitrary precision \cite{Verstraete2006}. 
For the long-range behavior of the system, however, this is not necessarily the case as correlations of MPS with finite bond dimension by construction  must decay purely exponentially at sufficiently long distances \cite{Fannes1992}. 

The question of how well MPS are able to reproduce correlations at long distances becomes particularly interesting in view of recent observations linking the minima of dispersion relations of elementary excitations of local, translationally invariant Hamiltonian with the rate of decay of momentum-filtered correlations in its ground state \cite{Zauner2014,Haegeman2014}.  Moreover, a novel interpretation of the transfer matrix obtained in the MPS algorithm was also proposed in Ref.~\onlinecite{Zauner2014}. Namely, that it reproduces the quantum transfer matrix in the Euclidean path-integral representation of the quantum state through a renormalisation group procedure equivalent to the seminal Wilson's numerical renormalisation group \cite{Wilson1975}. In that picture the physical spin is interpreted as an impurity and the MPS transfer matrix contains only the subset of degrees of freedom, out of exponentially many for the quantum transfer matrix, which are relevant for description of its correlations. This interpretation still  requires corroboration by explicit calculations, which we partially address in this article -- see also Ref. \onlinecite{Bal2015} in that context.

A closely related topic is that of the Finite Entanglement Scaling at the critical point \cite{Tagliacozzo2008,Pollmann2009,Vid2014} -- a numerically established fact that in an MPS approximation of the ground state of a critical system, there emerges a long-distance correlation length as an artifact of the finite bond dimension of the MPS. The state exhibits scaling behavior as a function of growing bond dimension, and it is even possible to extract the conformal information about the critical point by analyzing it \cite{Vid2014}. Nevertheless, it remains an interesting topic to get a better understanding of how Finite Entanglement Scaling arises.

In this article we study a particular example where such questions can be addressed analytically, albeit in the framework of MPS. 
To that end, in Sec. II, we show how to construct an exact MPS representation of the ground state of the XY model -- a prototypical spin model in one dimension -- with in principle exponentially diverging bond dimension.  As a corollary, we use this representation to illustrate how the Ornstein-Zernike form of the correlation function naturally emerges in this
(exact) case using the standard language for MPS.

Subsequently, in Sec. III, we show how to obtain an MPS representation with finite bond dimension from the one discussed above. We examine how this truncation -- which is also a part of many numerical algorithms -- affects the state, focusing mostly on the long distance correlations and on the spectrum of the transfer matrix.  In particular, in Sec. IIIA, we analyze the error in reproducing the correlation length in the gapped system.   In Sec. IIIB we examine the relation between our construction and the Finite Entanglement Scaling at the critical point, and find close similarities, allowing a geometric interpretation.  Finally, in Secs. IIIC and IIID, we take a comprehensive view on the spectrum of the transfer matrix, as well as the form factors for the correlation functions, providing direct  evidence in support of the impurity picture, as proposed in Ref.~\cite{Zauner2014}.

\section{The ground state of the XY model and its exact MPS representation} 

The $S=1/2$ XY model on a chain of $N$ spins is defined by the Hamiltonian
\small
\begin{equation} \label{eq:HXY}
   H_{XY} = -\sum_{n=1}^N\Br{\frac{1+\gamma}{2} \sigma_n^x \sigma_{n+1}^x+\frac{1-\gamma}{2} \sigma_n^y \sigma_{n+1}^y + g \sigma^z_n},
\end{equation}
\normalsize
where $\sigma^{x,y,z}_n$ are standard Pauli operators acting on site $n$, and periodic boundary conditions are assumed.
In order to construct an MPS representation of the ground state $|\Psi_{XY} \rangle$, we exploit the 1D-quantum to 2D-classical mapping (commonly used e.g. in the context of Quantum Monte Carlo \cite{MonteCarlo}, Corner Transfer Matrix DMRG \cite{CTM_DMRG}, or Bethe Ansatz at finite temperature \cite{BetheAnsatz}, to name just a few),   
and above all the original observation by Suzuki \cite{Suzuki1971} that $H_{XY}$ commutes and -- more importantly -- shares the ground state with an operator $V$ given by
\begin{eqnarray}  \label{eq:Vop} 
      &V = V_1^{\frac12} V_2 V_1^{\frac12}, \\
      &V_1 = \exp\Br{\overline K_1 \sum_{n=1}^N \sigma^z_n};  \ V_2 = \exp\Br{ K_2 \sum_{n=1}^N \sigma^x_n \sigma^x_{n+1}}, \nonumber
\end{eqnarray}
which appears naturally as the transfer matrix in the solution of the classical 2D Ising model \cite{RMP_LSM}.
We follow the notation of \cite{RMP_LSM} for convenience but in our case the parameters $\overline K_1$ and $K_2$, which we assume are non-negative, a priori do not have any specific physical interpretation.

For the sake of clarity we briefly reiterate the main steps of diagonalizing $V$. The subsequent use of a Jordan-Wigner transformation  
$\sigma^z_n = 1-2 c_n^\dagger c_n$, $\sigma^x_n + i \sigma^y_n = 2 c_n \prod_{m<n} \left( 1-2 c^\dagger_m c_m \right)$ 
[with fermionic annihilation operators $c_n$], a Fourier transform 
$c_n = e^{-i \pi/4} N^{-\frac12} \sum_k c_k e^{i k n}$, 
and a Bogolyubov transformation 
$c_k =\cos \theta_k \gamma_k - \sin \theta_k \gamma_{-k}^\dagger$ 
allows to rewrite $V$ as \cite{2DIsing_comment,normalization_comment}:
\begin{equation} \label{eq:Vdiag}
V = \exp \left[- \sum_k \epsilon_k \gamma_k^\dagger \gamma_{k} \right]=\exp \left[- H_V \right],
\end{equation}
where the single particle energies $\epsilon_k \ge 0$ are given by
$\cosh \epsilon_k = \cosh 2 \overline K_1 \cosh 2 K_2 - \cos k \sinh 2 \overline K_1 \sinh 2 K_2$,
and the Bogolyubov angles $\theta_k$ are determined as
\begin{equation}
\tan 2 \theta_k = \frac{\sin k  \sinh 2 K_2}{  \sinh 2 \overline K_1 \cosh 2 K_2 - \cos k \cosh 2 \overline K_1 \sinh 2 K_2  }.
\label{eq:BogolyubovV}
\end{equation}

\begin{figure}[t] 
\begin{center}
\includegraphics{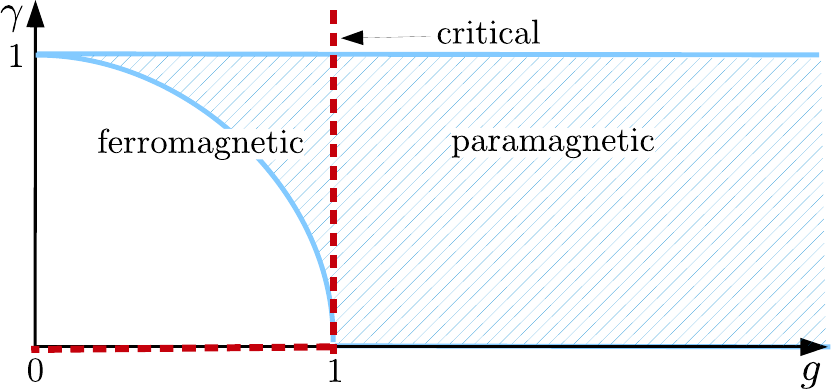}
\end{center}
\caption{(color online). Phase diagram of the XY model in Eq.~\eqref{eq:HXY}. Blue hatching (without boundaries) shows the range of parameters covered by the mapping in Eq.~\eqref{eq:XYmap}, where the boundaries are given by $\gamma=1(0)$  and $\gamma^2 + g^2 = 1$. Red dashed lines indicate critical lines.}
\label{fig:pd}
\end{figure}

The Hamiltonian of the XY model \eqref{eq:HXY} can be diagonalized following exactly the same steps as those for $V$, but with the Bogolyubov angles given by
\begin{equation}
\tan 2 \theta_k  = \frac{\gamma \sin k}  {g - \cos k}.
\label{eq:BogolyubovXY}
\end{equation}
Therefore, they can be simultaneously diagonalized if
\begin{equation} \label{eq:XYmap}
  g = \frac{\tanh 2 \overline K_1}{ \tanh 2 K_2}, \quad  \gamma = \frac{1}{\cosh 2 \overline K_1},  
\end{equation}
in which case the Bogolyubov angles in Eqs. \eqref{eq:BogolyubovV} and \eqref{eq:BogolyubovXY} match and $H_{XY}$ and $H_V$ commute. Subsequently we check that the ground state of one is also the ground state of the other.
The mapping above covers the part of the phase diagram of the XY model where $0<\gamma<1$ and $\gamma^2 + g^2 >1$ as shown in Fig.~\ref{fig:pd}. 

It is worth pointing out, that one could match the Bogolyubov angles also by considering complex $K_2 \to K_2 + \frac{i \pi} 4$, which, as follows from Eq.~\eqref{eq:XYmap}, would cover the incommensurate region of $\gamma^2 + g^2 <1$. However, while the Hamiltonians $H_{XY}$ and $H_{V}$ do commute, they only share the ground state if an additional condition of $\gamma^2 + g > 1$ is satisfied. For that reason, a more general approach is required to extend the mapping to the incommensurate region with oscillating correlation function and, in this article, we limit ourself only to the commensurate case of $\gamma^2 + g^2 >1$ and real $\overline K_1$ and $K_2$.

At the risk of stating the obvious, $V$ appears naturally also as a result of the second order Suzuki-Trotter expansion of (the exponent of) the quantum Ising model Hamiltonian $H_I = -\sum_{n}\Br{J \sigma_n^x \sigma_{n+1}^x+ \Gamma \sigma^z_n}$, in which case $\overline K_1 = dt \Gamma$ and $K_2 = dt J$, where $dt$ is a discrete Trotter step. Equation \eqref{eq:XYmap} precisely quantifies the Trotter errors for this specific, but commonly employed case, by showing that one approaches the ground state of the XY model as a result of them.


\subsection{MPS representation of the ground state}

Exponentials of operators of the form in Eq.~\eqref{eq:Vop} can be efficiently decomposed in terms of Matrix Product Operators (MPOs) with bond dimension $2$ \cite{Murg2010}:
\begin{equation} \label{eq:VMPO}
V =  \sum_{s_1,\dotsc, s_N=0}^1   \Tr \Br{C^{s_1} \dotsm C^{s_N}} X^{s_1}  \otimes \dotsm \otimes X^{s_N},
\end{equation}
up to normalization \cite{normalization_comment}, where
\begin{equation*}
\begin{split}
      C^0 = \sqrt2 \begin{pmatrix} \cosh K_2 & 0  \\ 0&   \sinh K_2 \end{pmatrix}, \;
      C^1 = \sqrt{\sinh 2 K_2} \begin{pmatrix} 0 & 1  \\ 1 & 0 \end{pmatrix}, \\
      X^0 = \sqrt2 \begin{pmatrix} \cosh K_1 & 0  \\ 0&   \sinh K_1 \end{pmatrix}, \;
      X^1 = \sqrt{\sinh 2 K_1} \begin{pmatrix} 0 & 1  \\ 1 & 0 \end{pmatrix}.
\end{split}
\end{equation*}
We refer to \cite{Murg2010} for details of the derivation. Following \cite{RMP_LSM}, we employ the convenient notation where $\overline K = -\frac12 \ln (\tanh K)$, and $\overline {\overline K} = K$.

\begin{figure}[t]
\begin{center}
\includegraphics[width=\columnwidth]{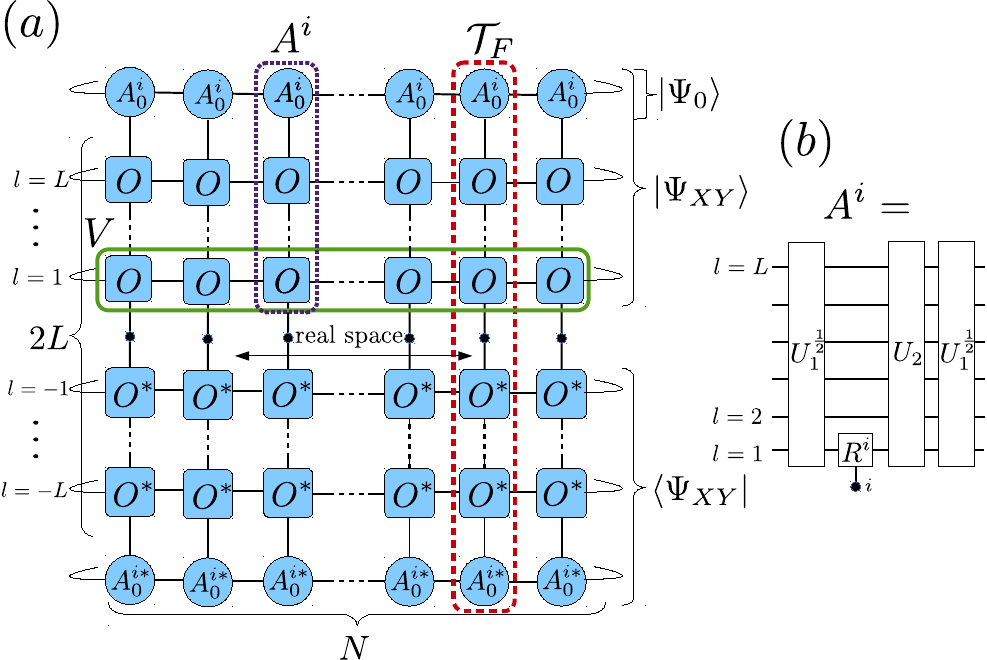}
\end{center}
\caption{(color online). (a) Decomposition of the ground state $|\Psi_{XY} \rangle$  into a two-dimensional tensor network. Rows represent operator $V$ in MPO form with local tensors $O$.
Half-columns constitute the MPS decomposition of $|\Psi_{XY} \rangle$ with MPS matrices $A^i$, while a full (infinite) column represents the (quantum) transfer matrix $\mathcal{T}_F$ at zero temperature. 
(b) MPS matrix $A_i$ as given by Eq.~\eqref{eq:A}.}
 \label{fig:network}
\end{figure}

By applying $V$ to some initial state $|\Psi_0\rangle$ $L$ times, the ground state $\vert \Psi_{XY}\rangle$ of the XY model is obtained in the limit $L \to \infty$ (pending normalization).
This is equivalent to performing imaginary time evolution with the Hamiltonian $H_V$ in Eq.~\eqref{eq:Vdiag}, where the effective imaginary time of evolution is proportional to $L$. 
This procedure is depicted in the top half of Fig.~\ref{fig:network}a, where each row represents a single operator $V$ in MPO form with local tensors $O = \sum_{s=0}^1 C^s \otimes X^s$.

Alternatively, one can look at this picture in the vertical direction, interpreting each column $A^i$ as an exact MPS representation of $\vert \Psi_{XY}\rangle$ with bond dimension $2^L$, that is,
$\vert \Psi_{XY}\rangle  = \sum_{i_1,\dotsc, i_N}   \Tr \Br{A^{i_1} \dotsm A^{i_N}} \vert i_1\dotsc i_N  \rangle = \vert \Psi(A)\rangle$.
Due to the symmetry between $C^s$ and $X^s$, which is apparent from Eq.~\eqref{eq:VMPO}, we can obtain $A^i$ simply by inverting the steps leading to the MPO decomposition of $V$.
The only additional complication comes from the boundaries of $A^{i}$, representing the physical (spin) degrees of freedom and the initial state $\vert \Psi_0 \rangle$, respectively.
After some algebra we obtain
\begin{eqnarray}  \label{eq:A} 
      &A^i = U_1^{\frac12} R^i U_2 U_1^{\frac12}, \\
      &U_1 = \exp\Br{\overline K_2 \sum_{l=1}^{L} \tau^z_{l}}; \ U_2 =  \exp\Br{ K_1 \sum_{l=1}^{L-1} \tau^x_{l} \tau^x_{l+1}}. \nonumber
\end{eqnarray}
Here $l =1,2, \dotsc, L$ labels the auxiliary degrees of freedom along the vertical direction and $\tau^{x,y,z}_l$ are Pauli operators acting on these. 
$R^i$ is a localized operator acting on auxiliary site $l=1$, with $R^0 = \sqrt{\cosh K_1}\unity$ and $R^1 = \sqrt{\sinh K_1} \, \tau^x_1$.  
For a graphical representation see Fig.~\ref{fig:network}b.
Notice that $R^i$ commutes with $U_2$, but not with $U_1$.

As an initial (top) state we use for convenience a product state 
$\vert \Psi_0 \rangle = \vert 0_1 0_2 \dotsc 0_N \rangle$ 
with $\sigma^z_n |0_n\rangle = |0_n\rangle$ fully polarized in the $+Z$ direction. 
This state is an eigenstate of the parity operator $P = \prod_{n=1}^N \sigma^z_n $ with eigenvalue $+1$. 
As $P$ commutes with $V$, the final state $\vert \Psi_{XY} \rangle$ has the same parity as $\vert \Psi_0 \rangle$. 
In particular, this means that in the ferromagnetic phase we consider the symmetric superposition of the two symmetry broken ground states.

From now on, for the rest of this article, we will assume that the system is in the thermodynamic limit $N\to\infty$.

\subsection{ Correlation functions}

We have cast the ground state of the XY model in an exact MPS form, where each MPS matrix $A^i$ has bond dimension $2^L$ and the limit $L \rightarrow \infty$ is assumed. 
However, before we address the question of finding an efficient MPS approximation with low bond dimension, it is illuminating to discuss the asymptotic behavior of the correlation functions in the exact case.

Following standard notation \cite{Verstraete2008,Schollwock2011} we define the MPS transfer matrix (see Fig.~\ref{fig:network}a) as 
\begin{equation}
\mathcal{T}_F  = \sum_{i=0}^1 \bar{A}^{i} \otimes A^i,
\end{equation}
which, up to the open boundary conditions and exchanging $K_1$ and $K_2$, has exactly the same form as $V$ in Eq.~\eqref{eq:Vop}.  We discuss this in more detail in the Appendix.
The subscript {\it F} denotes that this is the full transfer matrix of the exact MPS representation of the ground state, as opposed to the transfer matrix of the truncated MPS, which will be introduced in the next section.

We also define the (spin) operator transfer matrix  $\mathcal{T}_{F}^{\hat o} = \sum_{i,j=0}^1 \hat o_{i,j} \bar{A}^{i} \otimes A^j$, which obviously simplifies to the one above for $\hat o = \mathbb{1}$.  For $\hat o = \sigma^z$, which will be of most interest to us in this section, we find it convenient to express it as
$\mathcal{T}_{F}^{\sigma^z} = \mathcal{Q}_{z} \mathcal{T}_F$, where $\mathcal{Q}_z =  \exp \Br{-2 K_1 \tau_{-1} \tau_{1}}$ and $\tau_l = \cosh \overline K_2 \tau^x_l + i \sinh \overline K_2 \tau^y_l $. This way the effect of $\sigma^z$ in the spin transfer matrix is encoded in an operator which is localized in the virtual direction at sites with $l = \pm 1$, see Fig.  \ref{fig:network}a.  Note however that some care is needed here as $\mathcal{Q}_z$ is not hermitian and does not commute with $\mathcal{T}_F$.
The static correlation function is then calculated as $\langle \hat o_0 \hat o_R\rangle = \Tr\left(\mathcal{T}_F^{\hat o} \cdot \mathcal{T}_F^{R-1}  \cdot \mathcal{T}_F^{\hat o}  \cdot \mathcal{T}_F^{N-R-1} \right)/\Tr \left(\mathcal{T}_F^N \right)$. 

For illustrative purposes, we further consider only the connected correlation function
\begin{equation} \label{eq:correlations}
C_{zz}(R)= \langle \sigma^z_0 \sigma^z_R \rangle - \langle \sigma^z \rangle^2 =  \sum_{\alpha \neq \emptyset} f_\alpha^{zz} e^{ - E_\alpha R},
\end{equation}
where the second equality is valid in the thermodynamic limit when $\mathcal{T}_F^{N-R-1}$ projects onto the ground space of $\mathcal{T}_F$ -- which is hermitian by construction.
Here, we have defined the form factors as
$f_\alpha^{zz} = ( \varphi_\emptyset \vert \mathcal{Q}_z  \vert \varphi_\alpha ) ( \varphi_\alpha \vert \mathcal{Q}_z  \vert \varphi_\emptyset )$
using the localized operator  $\mathcal{Q}_z$ for simplicity.  $|\varphi_\alpha )$ are the eigenvectors of the transfer matrix ${\mathcal{T}_F}$ to eigenvalues $e^{- E_\alpha }$ and $|\varphi_\emptyset )$ is the dominant one. Below we will only consider the case where $|\varphi_\emptyset )$ is unique. Otherwise, e.g. in the ferromagnetic phase, 
the definition of form factor has to be generalized to include properly normalized sum over all dominant eigenvectors.

${\mathcal{T}_F}$  can be diagonalized by mapping onto a free-fermionic system as
\begin{equation}
 \mathcal{T}_F = \exp \Br{-\sum_{k}\epsilon_k^{\mathcal{T}_F} a_k^\dagger a_k},
 \label{eq:diagonalTF}
\end{equation}
where $a_k$ are fermionic annihilation operators. For simplicity, in this section, we approximate $\mathcal{T}_F$ by using periodic boundary conditions, which does not affect the results. It is then diagonalized following the same steps as for $V$ in Eqs. \eqref{eq:Vop}-\eqref{eq:Vdiag}. 
In the limit $L\rightarrow \infty$ the spectrum of the transfer matrix consists of continuous bands as the states $\vert\varphi_{\alpha \neq \emptyset})$ are obtained from the vacuum $\vert \varphi_\emptyset )$ by exciting free-fermionic quasiparticles and  $E_\alpha$ follows from summing up the corresponding single particle energies $\epsilon_k^{\mathcal{T}_F}$, where
\begin{equation}
\cosh \epsilon_k^{\mathcal{T}_F} = \cosh 2 \overline K_2 \cosh 2 K_1 - \cos k \sinh 2 \overline K_2 \sinh 2 K_1.
\label{eq:dispersionTF}
\end{equation}

Now, in order to obtain the leading asymptotic of the correlation function, it is sufficient to know the dispersion around the minimum of the lowest relevant band -- i.e. for which the form factors $f_\alpha^{zz}$ are nonzero -- and the scaling of those form factors.
In the case of $C_{zz}(R)$ the only nonzero form factor contributions come from $\alpha = \{k_1,k_2 \}$, that is, where two quasiparticles with momenta $k_1$ and $k_2$ are excited.
Notice that form factors corresponding to the lowest single particle band $\alpha=\{k_1\}$ vanish, since both $\mathcal{T}_{F}$ and $\mathcal{Q}_z$ conserve parity. Below, we consider two cases:

I) Critical point for $g=1$, in which case $K_1 = \overline K_2$. Expanding around the minimum of $ \epsilon_k^{\mathcal{T}_F}$ at $k=0$ we obtain
\begin{equation}
      \epsilon^{\mathcal{T}_F}_k \simeq a_c |k|; \ \  f^{zz}_{k_1,k_2}  \simeq b_c \frac{\pi^2}{L^2}, \ \ \mathrm{ for }  \ k_1 \cdot k_2 <0,
\end{equation}
with the coefficients $a_c = \sinh 2 K_1$ and $b_c = a_c^2 / \pi^2$.  
Now, for large $R$ the correlation function behaves asymptotically as
\begin{equation}
C_{zz}(R)  \approx \sum_{k_1>0;k_2 < 0} \frac{\pi^2}{L^2} b_c   e^{-R ~ a_c (|k_1| + |k_2|)}.
\end{equation}
In the limit of $L \to \infty$ we treat $k_{1,2}$ as continuous variables with $dk_{1,2} = \frac{2\pi}{2L}$ and we have
\begin{equation}
C_{zz}(R)  \approx \iint_{0}^{\infty} dk_1 dk_2 b_c e^{-R ~ a_c (k_1+ k_2)},
\label{Czzint1}
\end{equation}
where we can extend the limits of integration to $+ \infty$ for large $a_c R$. 
Rescaling the variables in the integral leads to the algebraic dependence on $R$,
\begin{equation}
C_{zz}(R)  \approx \frac{b_c }{a_c^2 } \frac{1}{R^2} =\frac{1 }{\pi^2} \frac{1}{R^2}, 
\end{equation}
where we recover the classic result by Barouch and McCoy \cite{Barouch1971}.
 
II) Paramagnetic phase for $g>1$. The dispersion relation exhibits qualitatively different behavior around its minimum when compared to the critical case discussed above. Expanding around $k=0$ we obtain
\begin{equation}
      \epsilon^{\mathcal{T}_F}_k \simeq \Delta +  a_p  k^2; \ \  f^{zz}_{k_1,k_2}  \simeq b_p  (k_1-k_2)^2 \frac{\pi^2}{L^2} ,
\end{equation}
with the gap $\Delta=2 \vert \overline K_2 - K_1\vert $ and the coefficients 
$a_p = \sinh(2 \overline K_2)\sinh(2 K_1) / 2 \sinh(2 \vert \overline K_2 - K_1\vert)$ 
and 
$b_p = a_p^2 / \pi^2$.
Now, for large $R$ and in the continuous limit of $L\to \infty$ the correlation function behaves asymptotically as
\begin{equation}
C_{zz}(R)  \approx e^{-2 \Delta R }  \iint_{-\infty}^{\infty} \frac12 dk_1 dk_2 b_p (k_1 -k_2)^2 e^{-R a_p (k_1^2 + k_2^2)},
\label{Czzint2}
\end{equation}
where we can extend the limits of integration to $\pm \infty$ for large $a_p R$. 
Naturally, we recognize the correlation length as $\xi = \Delta^{-1}$, which is the slowest possible decay resulting from $\mathcal{T}^F$. 
Performing integrals -- or just rescaling variables to extract $R$ from the integrals -- yields the leading algebraic dependence on $R$,
\begin{equation}
C_{zz}(R)  \approx \frac{b_p \pi }{2 a_p^2 } \frac{1}{R^2}e^{-2 R/\xi }=\frac{1 }{2 \pi } \frac{1}{R^2}e^{-2 R/\xi }, 
\label{Czzgap}
\end{equation}
again in agreement with \cite{Barouch1971}. It is worth pointing out that this expansion is valid for $R \gg \xi$, what can be seen from the size of $a_p R$. For smaller $R$, when the exponential does not suppress other terms, the correlation function would behave similarly as at the critical point. While this remark is not so important for ZZ correlation here, other correlation functions might have other exponents of the algebraic part at the critical point and away from it \cite{Barouch1971}.

To summarize, notice that on the one hand the gap of the transfer matrix sets the correlation length while on the other hand, the full (low energy part of the) continuous band contributes {\it equally} (in the sense of form factors being proportional to $dk$) to form the algebraic part as a consequence of the nonflat dispersion relation.
The value of the exponent in this algebraic behavior is determined by a combination of the dispersions of the leading eigenvalues of the transfer matrix, and of the corresponding form factors, as well as symmetries. For instance in the case of $C_{zz}$ discussed above, the conserved parity symmetry modifies this exponent due to the double integrals appearing in Eqs. \eqref{Czzint1} and \eqref{Czzint2}. This is also accompanied by halving the correlation length in Eq.~\eqref{Czzgap} comparing to the slowest possible decay suggested by the transfer matrix. We also note that, at the critical point, this picture might be more complicated as contributions from many bands are relevant in some cases; we refer to the Appendix for additional details.

Such an asymptotic behavior of the correlation function is often labeled in literature as the Ornstein-Zernike form \cite{Ornstein1914,Kennedy1991}, especially in context of systems significantly away from the critical points. For 1d quantum systems, it is typically expected that the exponent in the algebraic behavior is equal to $\frac12$, namely $C(R) \sim R^{-1/2} e^{-r/\xi}$.  Notice that this form of asymptotic behavior would indeed emerge in our treatment in the most simple case when single-particle form factors are nonzero (they might be zero for example as a result of some symmetries of local observables) and in the leading order independent on $k$: $f_k \simeq \mathrm{const} \cdot dk$, and the dispersion relation around the minimum is smooth and quadratic $\epsilon^{\mathcal{T}_F}_k \simeq \Delta +  a  (k-k_{min})^2$. This happens e.g. for the XX correlation function in the paramagnetic phase.
We discuss this further in the Appendix, where we numerically obtain the behavior of the form factors for other correlation functions in various phases.


\section{Characterizing the efficient MPS approximation}

The results above were obtained analytically in the limit of exponentially diverging bond dimension $D$, where $D=2^L$ with $L\to \infty$. 
However, above all, MPS serve as a class of variational states that lie at the heart of many numerical techniques, where only modest bond dimensions are feasible. It is therefore important to understand what information about the quasi-exact state is retained after truncating to an efficient MPS approximation with finite bond dimension $D$, i.e. $ \lvert \Psi (A) \rangle \simeq \lvert {\Psi} (\tilde A) \rangle$, with $\tilde A^i$ matrices $\in\mathbb{C}^{D\times D}$.

 We follow the standard MPS truncation procedures for infinite, translationally invariant systems \cite{Orus2008}. It is based on finding the Schmidt decomposition of the state along a single cut and retaining only the dominant Schmidt values. While such a procedure would be optimal for truncation at a single bond, in the infinite system it is performed at all sites simultaneously \cite{PBC_comment}.

This is then equivalent to finding the reduced density operator of the MPS $|\Psi(A)\rangle$ on a half-infinite chain and finding its diagonal basis in which we truncate by keeping only the dominant eigenvalues.
For this particular case, where the transfer matrix $\mathcal{T}_F$ is Hermitian and its left and right dominant eigenvector are both given by $|\varphi_\emptyset)$, the physical density operator of the half infinite chain shares the spectrum with the reduced density matrix $\rho$ of $|\varphi_\emptyset)$ with support on the $L$ site auxiliary system with $l>0$ (see Fig.~\ref{fig:network}).

We briefly outline the main steps of the procedure below and refer to the Appendix for details. We map the full transfer matrix onto a system of free fermions and, following Ref.~\onlinecite{Abraham1971}, describe the transfer matrix using {\it the transformation matrix}, which describes the transformation of fermionic operators under the similarity transformation given by $\mathcal{T}_F$. Since part of the operations has to be performed numerically, we keep $L$ large but finite. Bringing the transformation matrix into the canonical form allows to find the Bogolyubov transformation which diagonalizes the transfer matrix and $|\psi_\emptyset)$ is the vacuum state in that basis. We describe the reduced density matrices in a standard way \cite{Peschel2003,Peschel2009} by using the two-point correlation matrix. Subsequently, it can then be expressed as 
$\rho = \frac{1}{Z} \exp \left(- 2 \sum_{m=1}^L  \delta_m f^\dagger_m f_m \right)$, 
where $f_m$ are fermionic annihilation operators and $\delta_m>0$ is the entanglement spectrum arranged in ascending order.  $f_m$ and $\delta_m$ are obtained by finding the Bogolyubov transformation which brings the correlation matrix into its canonical form.  Efficiently truncating to an MPS with bond dimension $D=2^{\chi}$ is now obtained by keeping only the first $\chi$ (most relevant) \textit{fermionic modes} of $\rho$.
This amounts to the projection
\begin{equation}
\label{eq:Anew}
\tilde A^i =  ( 0_{\chi +1} 0_{\chi +2} \dotsc |A^i | 0_{\chi +1} 0_{\chi +2} \dotsc ),
\end{equation}
where $f_m |0_m) =0 $.
 
We point out that this procedure is {\it not} fully equivalent to just keeping the $D$ largest singular values of $\rho$ as it additionally retains the free-fermionic structure of the problem, as
the reduced density matrix of the truncated state -- and consequently its Schmidt values -- have the form
\begin{equation}
\label{eq:newrho}
\tilde \rho = \frac1Z \exp \left(- 2 \sum_{m=1}^\chi \delta_m f^\dagger_m f_m \right).
\end{equation} 
This means that some care is needed when we compare the bond dimensions (and states) obtained with our procedure with those from conventional numerical MPS methods. For instance, $\tilde \rho$ is going to contain some very small Schmidt values together with the dominant one, especially for larger $\chi$.
As a trade-off, the structure we retain allows for a clean description and interpretation of the spectra, which would be nearly impossible using the standard numerical approach.  

Finally, as described in the Appendix, we obtain the transformation matrix describing the transfer matrix generated by 
$|\Psi (\tilde A)\rangle$, i.e. $\tilde{ \mathcal{T}} = \sum_{i=0}^1 \bar{\tilde{A}}^{i} \otimes \tilde A^i$. By finding the Bogolyubov transformation which brings it into the canonical form we diagonalize the truncated transfer matrix as
\begin{equation} \label{eq:teps}
\tilde{ \mathcal{T}} = \exp\left[- \sum_{m=1}^{2\chi} \tilde \epsilon_m d^\dagger_m d_m \right],
\end{equation}
where the spectrum  is determined by single particle energies $\tilde \epsilon_m >0 $ arranged in ascending order. 
For the remainder of this article we will mostly focus on this spectrum.


\subsection{Dominant modes in the gapped system}
\begin{figure}[t] 
\begin{center}
\includegraphics[width=\columnwidth]{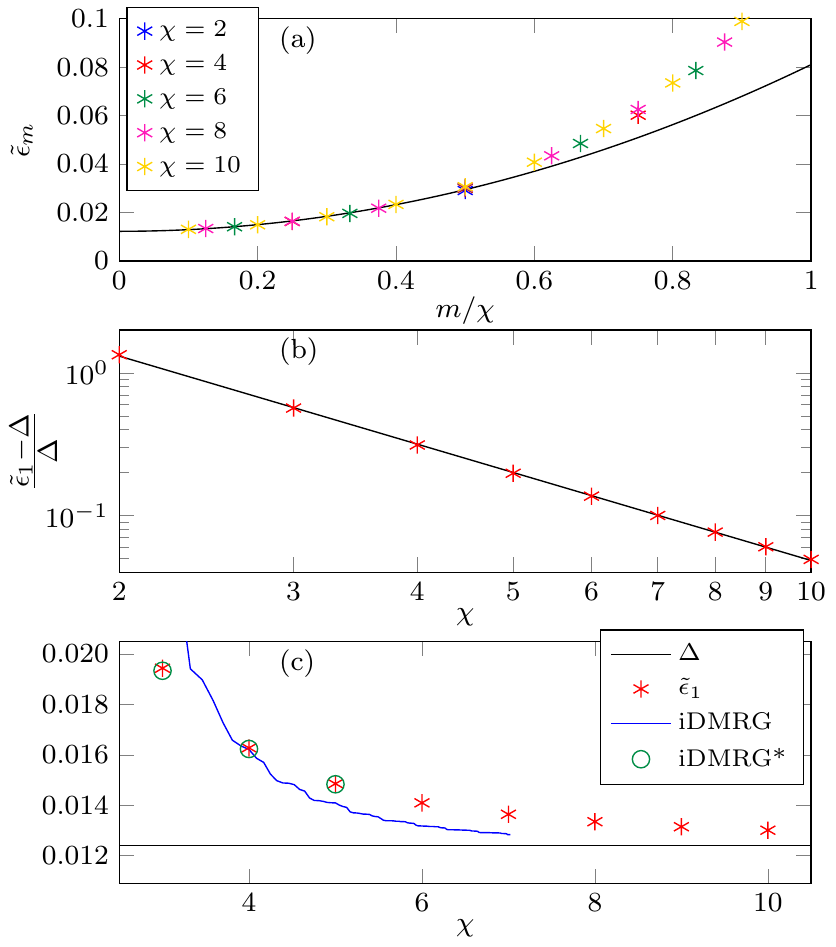}
\end{center}
\caption{(color online) XY model in the gapped phase with $g=1.01$ and $\gamma = 0.8$. 
(a) Single particle energies $\tilde \epsilon_m$ of Eq.~\eqref{eq:teps} for different $\chi$.  For rescaled index $m/\chi$, the data points collapse onto a single curve. The solid line is a quadratic fit to the first 4 points with $\chi=10$.
(b) Relative error of the correlation length $(\tilde \epsilon_1-\Delta)/\Delta$ as a function of $\chi$. It shows power-law behavior $\log \big[(\tilde \epsilon_1 - \Delta)/\Delta\big] = -2.0518 \log \chi +1.6944$ (green line).  
(c) Comparison with iDMRG calculations. The black line represents the exact gap $\Delta$ and stars represent the gap $\tilde \epsilon_1$ of the finite $\chi$ approximation of the transfer matrix. The blue line shows (minus log of) the transfer matrix gap obtained from an iDMRG ground state approximation, plotted as a function of $\log_2(D)$. Finally, circles represent data obtained from truncating iDMRG with $D'=120$ down to a smaller bond dimension while only selecting Schmidt states preserving the free-fermionic structure in Eq.~\eqref{eq:newrho}. }
\label{fig:numgap}
\end{figure}

Firstly, we discuss the non-critical case  focusing on the dominant eigenvalues of the transfer matrix $\tilde{\mathcal{T}}$.
To that end, we simulate the XY model for a particular set of parameters $g=1.01$ and $\gamma=0.8$ in the paramagnetic phase.   
We show the resulting single particle energies $\tilde \epsilon_m$ for several different values of $\chi$ in Fig \ref{fig:numgap}a.

Notably, we observe that the low energy part of the spectrum collapses onto a single curve when the index $m=1,2 \ldots, 2 \chi$ is rescaled by $\chi$. 
Moreover, the lowest part of the spectrum shows quadratic behavior in $m$, which corroborates a similar observation made in Ref.~\onlinecite{Zauner2014} for the case of conventional MPS calculations and part of the transfer matrix spectrum most relevant for the XX correlation function, see Fig.~9 therein.

In other words the results are consistent with the scaling of the form $\tilde \epsilon_m -\Delta \simeq a (m/\chi)^2$, at least for the first few $m$'s.  This universal quadratic behavior implies that the gap of the truncated transfer matrix $\tilde \epsilon_1$ should be shifted from the value of the true gap $\Delta$, and approach it as a power law in $\chi$ with exponent equal 2. 
That is indeed observed in Fig \ref{fig:numgap}b where we show the relative error in the gap (i.e. the inverse of the correlation length) for increasing $\chi$. We fit
\begin{equation}
(\tilde \epsilon_1-\Delta)/\Delta \simeq p_1 \chi^{-p_2},
\label{eq:eps1chi}
\end{equation}
with $p_2 \simeq 2.0518$ close to 2.  Notice, that for this particular set of parameters, even for $\chi=10$ the correlation length obtained from the free-fermionic MPS still underestimates the exact value by $\simeq 5 \%$, where the exact gap of the full transfer matrix is given as $\Delta = \epsilon_{k=0}^{\mathcal{T}_F}  = 0.0124035 \dotsc$
Those observations also hold for other values of the magnetic field, where $p_2$ remains close to 2 and $p_1$ grows slowly when approaching the critical point.

It is apparent that a finite value of $\chi$ results both in an underestimation of the correlation length and the breakdown of the asymptotic algebraic dependence of the correlation function on $R$ above some length scale dictated by $\chi$. 
Correspondingly, by increasing $\chi$, the MPS is able to better reconstruct the low energy part of the continuous band which is responsible for the asymptotic algebraic part of the exact correlation function.

The neat behavior of the gap observed above should be contrasted with the results from the conventional MPS calculations using iDMRG \cite{DMRG,McCulloch2008}, which we plot in Fig.~\ref{fig:numgap}c.  First of all, we obtain a very good match between free-fermionic results and iDMRG, provided the correct $2^{\chi}$ Schmidt-states are selected from an MPS obtained from iDMRG with initially larger bond dimension (labeled as iDMRG* \cite{iDMRG*}) corroborating our procedure.

Without preserving the structure of Eq.~\eqref{eq:newrho}, but rather just keeping the largest Schmidt values during truncation, standard iDMRG is able to approach the exact gap $\Delta$ faster with increasing $D$, but in a rather irregular, step-wise way. Remarkably, we can use the observations made for free-fermions above to estimate this behavior.

We employ the fact that in the paramagnetic phase considered here the single particle Schmidt spectrum in Eq.~\eqref{eq:newrho}  has a simple form $\delta_m = (2m-1) \delta_1$ \cite{Peschel2009}. We define $D_{\chi}+1$ as the index of the largest Schmidt value {\it not} reproduced by Eq.~\eqref{eq:newrho} for given $\chi$, and use the form above to calculate the value of $D_\chi$. We can expect that the error of the gap obtained with iDMRG with bond dimension $D_{\chi}$ should be lower-bounded by the error of $\tilde \epsilon_1 (\chi)$ from our fermionic procedure, as the second one represents the state containing additional information coming from some smaller Schmidt values as well (numerics confirms this). $D_{\chi}$ grows quickly with $\chi$, and locally, for $\chi \approx 10$ we numerically see the scaling $D_{\chi} \sim \chi^{2.5}$, $(D_{\chi=10} = 63)$. By substituting this into Eq.~\eqref{eq:eps1chi} with $p_2 =2$ we expect that the error of the gap obtained with iDMRG with bond dimension $D$ should be vanishing slower than $D^{-0.80}$, at least for $D$ of the order of hundred. The fit (not plotted) to the data in Fig.~\ref{fig:numgap}c and $60<D<130$ shows that the error is shrinking on average as $D^{-0.79}$, remarkably close to our prediction.

Those results illustrate that one has to  take some care when extracting the correlation length directly from the transfer matrix, as the error can be vanishing slowly (and increasingly so) with growing bond dimension.


\begin{figure}[t] 
\begin{center}
\includegraphics[width=\columnwidth]{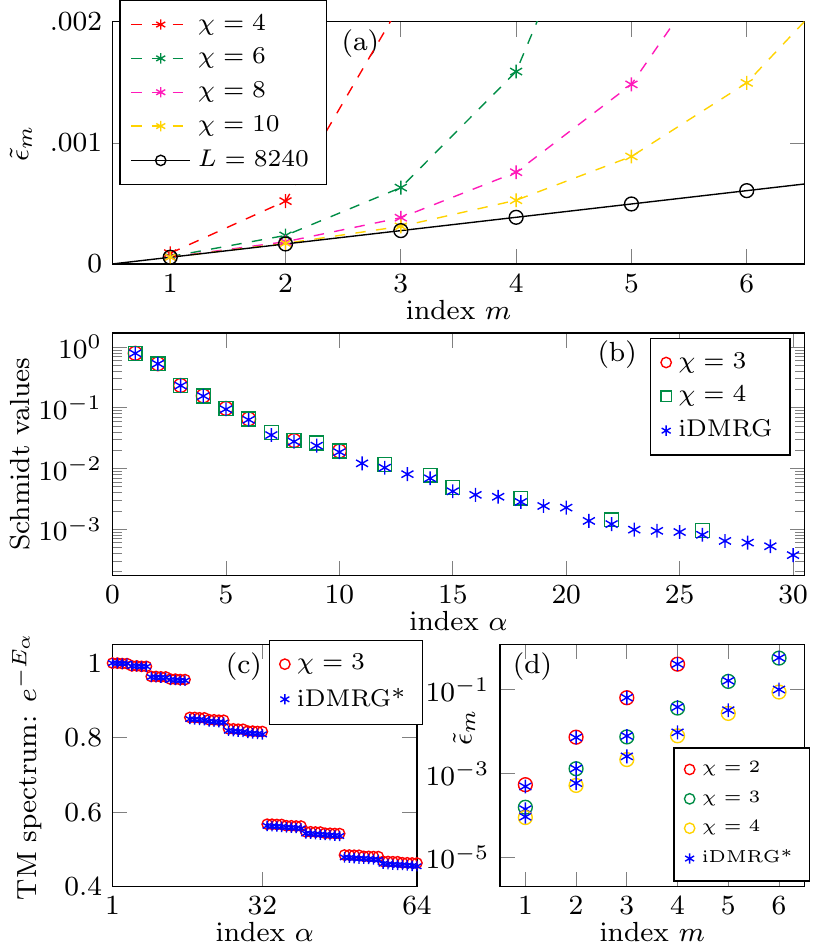}
\end{center}
\caption{(color online). XY model at a critical point with $g=1$ and $\gamma = 0.5$. (a) Smallest single particle energies $\tilde \epsilon_m$ for different $\chi$ quickly approach the exact smallest values from $L = 8240$. (b) Schmidt values obtained with iDMRG with $D'=70$ and corresponding Schmidt values obtained from $L=8240$ and truncating to $\chi=3,4$. 
(c) Comparison of the transfer matrix spectrum obtained for $L=8240$ truncated to $\chi=3$, and the spectrum from truncating iDMRG results with $D'=70$ down to $D=2^{3}=8$ \cite{iDMRG*}.
(d) Comparison of single particle energies in Eq. \eqref{eq:teps} obtained from truncating from $L=8240$ and iDMRG with $D'=70$, see text for details.
}
\label{fig:num_cri}
\end{figure}

\subsection{Dominant modes in the critical system and Finite Entanglement Scaling}

Secondly, we consider the gapless case, where we simulate the XY model for a particular set of critical parameters $g=1$ and $\gamma=0.5$. 
Contrary to the gapped system, the energy gap of $V$ is vanishing here and the actual ground state cannot be reached for any finite value of $L$.

The dominant part of the spectrum of the full transfer matrix $\mathcal{T}_F$ behaves as $\epsilon_k^{\mathcal{T}_F} \simeq a_c k$, where
the quasi-momenta take -- necessarily discrete -- values $k=\frac{\pi}{2L} (1,3,5,7,\dotsc)$, which are universal for systems with open boundary conditions.
This is shown in Fig.~\ref{fig:num_cri}a  for $L=8240$, together with the dominant $\tilde \epsilon_m$  from Eq. \eqref{eq:teps} obtained from truncation for several different $\chi$.  We observe that the $\tilde \epsilon_m$ reproduce the discrete low energy structure for finite $L$ increasing well with growing $\chi$.
In other words, even for relatively small $\chi$ the gap of the transfer matrix is effectively determined by $L$, which in turn is proportional to the imaginary time evolution of the initial state with Hamiltonian $H_V$, see Fig.~\ref{fig:network}.

On the other hand, conventional MPS calculations, where we use iDMRG \cite{DMRG,McCulloch2008}, try to approximate the critical ground state by the best possible state with finite correlation length \cite{Vid2014,Tagliacozzo2008,Pollmann2009,Pirvu2012}. 
A priori there is no reason to expect a good match between these two approximations (iDMRG and free-fermionic MPS with finite $L$), we however obtain a surprisingly good agreement nonetheless. To that end  we analyze the state obtained with iDMRG and bond dimension  $D'=70$ and compare it with data for $L=8240$. The value of $L$ is obtained here by matching the ratio of the first two Schmidt values, corresponding to $\delta_1$ in Eq.~\eqref{eq:newrho}, to the one obtained from iDMRG. 

In Fig.~\ref{fig:num_cri}b we plot the Schmidt values from iDMRG and the corresponding values for $\chi=3$ and $\chi=4$. Most importantly, even for iDMRG at the critical point, we are able to identify groups of Schmidt values corresponding  to the underlying free-fermionic structure given by Eq.~\eqref{eq:newrho}.  That such a structure is visible for the non-critical system is to be expected, however for the critical case there is a priori no reason for a numerical algorithm not to break it. 
Even more, the spectra obtained with iDMRG and with finite $L$ coincide rather well (i.e. corresponding  values of $\delta_m$ in Eq.~\eqref{eq:newrho} are  matching).

We further corroborate this by comparing the spectra of the transfer matrix  ($D^2=2^{2\chi}$ eigenvalues) for several values of $\chi$ and the corresponding iDMRG* \cite{iDMRG*}. 
As can be seen in Fig.~\ref{fig:num_cri}c for $\chi=3$, picking the correct Schmidt values results in a clear structure of the transfer matrix spectrum which is consistent with the one given by Eq.~\eqref{eq:teps} for finite $L$.
The structure of the transfer matrix in Fig.~\ref{fig:num_cri}c allows us to compute single particle energies corresponding to the ones in Eq.~\eqref{eq:teps} directly from iDMRG* and compare them with the ones obtained with free-fermions for finite $L$. 
The results are shown in Fig.~\ref{fig:num_cri}d  where the match for $\chi = 2,3,4$ is remarkably good. It is worth stressing here one more time, that all the points labeled iDMRG* in Fig.~\ref{fig:num_cri} are acquired  from the same initial state obtained from iDMRG with $D'=70$, which was subsequently truncated down by picking the correct Schmidt values.

The obtained results allow us to conclude that the state obtained with iDMRG contains the structure which is fully consistent with a free-fermionic theory on a strip of finite width with open boundary conditions -- for similar observation in finite system where the exact ground state can be reached due to the finite size effects, see \cite{Lauchli2013}. This provides a strong hint that so called Finite Entanglement Scaling \cite{Vid2014,Tagliacozzo2008,Pollmann2009,Pirvu2012} -- scaling observed while simulating the (conformally invariant) critical theory using MPS with finite $D$ -- can be interpreted in a geometric way. This cannot be seen that easily when looking directly at iDMRG and the ratios of the dominant eigenvalues of the transfer matrix (cf. \cite{Vid2014}), since the ratios are both highly susceptible to further truncations and even then, for a given state and $\chi =3,4$ they are still far from expected values of $(1,3,5,\dotsc)$ on a strip, see Fig.~\ref{fig:num_cri}a.

\begin{figure}[t] 
\begin{center}
\includegraphics[width=\columnwidth]{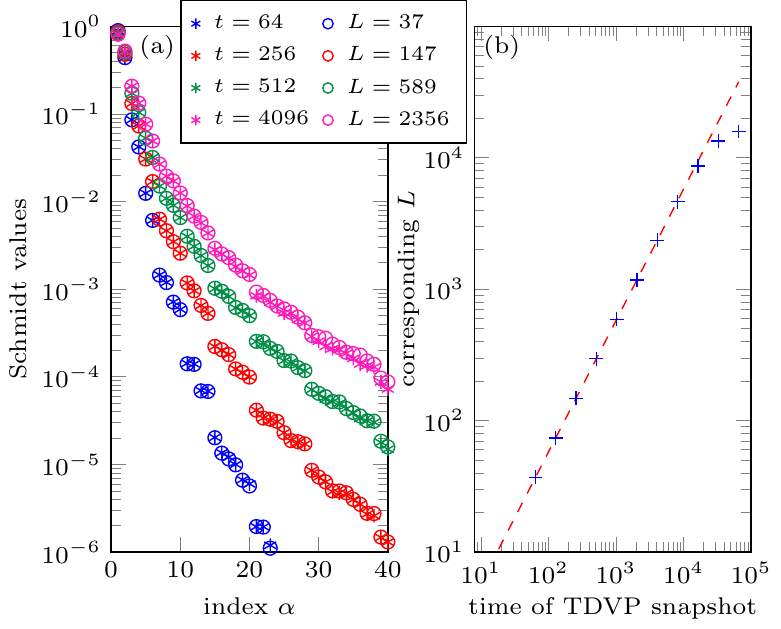}
\end{center}
\caption{(color online). XY model at a critical point with $g=1$ and $\gamma = 0.5$. (a) Dominant Schmidt values at snapshots of the imaginary time ($t$) evolution with the TDVP algorithm with $D'=100$ and corresponding Schmidt values from our free fermionic procedure with finite $L$. (b) For a large window of intermediate time the values of $L$ are proportional to $t$ (plotted in a log-log scale for convenience).}
\label{fig:TDVPvsff}
\end{figure}

However, while we were able to find a value of $L$ in a free-fermionic MPS which is a good match to a particular MPS obtained from iDMRG with given bond dimension $D'$, the above analysis does not provide any hint why, for given bond dimension $D'$, iDMRG should yield an MPS approximation corresponding to an effective imaginary time evolution of the system up to some finite imaginary time proportional to $L$.  Or equivalently, how to choose $D'$ for an iDMRG calculation to reproduce results from a finite $L$ free-fermionic calculation. 

In order to shred more light on this, we performed similar analysis for the TDVP algorithm \cite{Haegeman2011,Haegeman2014TDVP}  which is based on the imaginary time evolution. Likewise, we find a reasonable agreement between the state obtained with TDVP and our free-fermionic construction for some finite value of $L$,  provided that TDVP was initialized with the spin polarized state with even parity (all spins pointing in +Z direction). Then, the TDVP algorithm does not manifestly break the parity symmetry and along the evolution $\langle \sigma_x\rangle = \langle \sigma_y\rangle \approx 0$ (this is also the case for the state obtained with iDMRG above).  For TDVP initialized with the random state no good match with our construction could be found. 

In Fig. \ref{fig:TDVPvsff}a we show the dominant Schmidt values taken from a {\it single} run of the TDVP algorithm with $D'=100$, where we looked at the snapshots at different values of the imaginary time $t$.  We compare them with the corresponding Schmidt values from our free fermionic procedure with finite $L$, where the values of $L$ were obtained by matching the ratios of the two largest Schmidt values. For intermediate values of time the match is very good.  We also observe that it is getting considerably worse for large enough values of $t$ (around $t\sim 10^4$ in our case), where the energy of the TDVP state is effectively saturating at $10^{-10}$ above the exact ground state energy.

Not surprisingly, for a wide range of intermediate times, the values of $L$ are proportional to $t$,  see \ref{fig:TDVPvsff}b.
This suggest that instead of the standard approach based on analyzing the converged states for various values of D', it would be more natural to extract the conformal information about the ground state just from the snapshots of single run of the TDVP algorithm for fixed value of $D'$.

\subsection{Truncation as effective description of impurity}

Finally, while the previous two sections were focusing on the dominant part of the transfer matrix, we take a more comprehensive view on the spectrum here, valid both for the critical and gapped case alike. 

\begin{figure} [t]
	\begin{center}
 	 \includegraphics{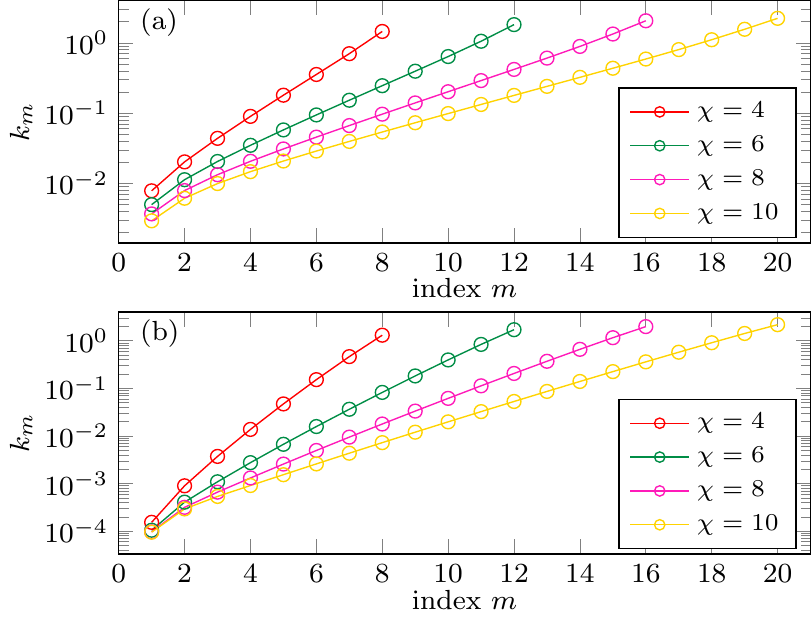}
	\end{center}
	\caption{(color online). Momenta $k_m$ corresponding to the single particle energies $\tilde \epsilon_m$ of the truncated transfer matrix for several  $\chi$. Data for (a) the gapped system with $g=1.01$ and $\gamma=0.8$, (b) critical system with $g=1$, $\gamma=0.5$ and $L=8240$. In both cases we find a clear linear dependence $\log k_m = m \log \lambda + \mathrm{const}$, valid for all but few smallest m's (i.e. smallest $\tilde \epsilon_m$).  } 
	\label{fig:5}
\end{figure}

To that end, we define discrete momenta $k_m$ corresponding to the spectrum of the truncated transfer matrix $\tilde \epsilon_m$ through the relation
\begin{equation}
\tilde \epsilon_m = \epsilon^{\mathcal{T}_F}(k_m).
\end{equation}
Above, $\epsilon^{\mathcal{T}_F}$ is the dispersion relation of the full transfer matrix, given by Eq. \eqref{eq:dispersionTF}, where the momenta take continuous (in the limit $L \to \infty$) values $k \in (0,\pi)$.
We show the resulting $k_m$, both for the gapped and critical cases, in Fig.~\ref{fig:5}. Most importantly, as can be seen in that plot, the value of momenta which are effectively selected during the truncation procedure satisfy  the relation
\begin{equation}
k_m \sim \lambda^m,
\label{eq:kmlambda}
\end{equation}
for all but the first few $m$'s.

In Ref.~\onlinecite{Zauner2014}, it was proposed that one can understand the transfer matrix obtained in the MPS algorithm as resulting from a renormalisation group procedure applied to the (full) quantum transfer matrix in the imaginary time direction. More precisely, the physical spin at $l=0$ in the virtual direction in Fig.~\ref{fig:network} plays a distinguished role in  
this tensor network as physical operators are applied there during the calculation of expectation values. It can then be interpreted as an impurity in the two-dimensional tensor network and the degrees of freedom relevant for the description of impurity -- and at the same time the concise description of the state in the MPS algorithm -- emerge as a result of application of Wilson's numerical renormalisation group (NRG) \cite{Wilson1975} to the quantum transfer matrix along the virtual (imaginary time) direction.

\begin{figure}[t]
\begin{center}
\includegraphics[width=0.8\columnwidth]{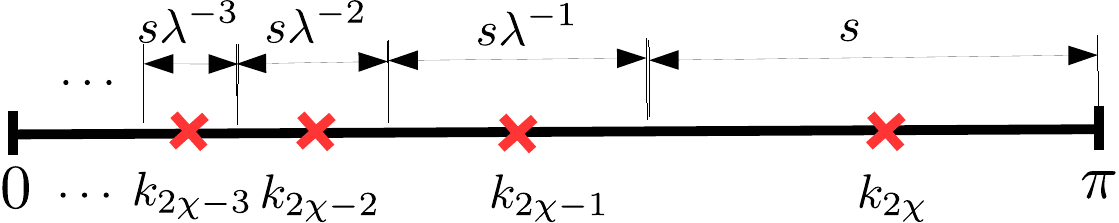}
\end{center}
\caption{(color online). Illustration of Wilson's renormalisation group in the momentum space. Each momentum window, logarithmically shrinking toward $k=0$, is represented by single effective mode.  This is also visible in Fig.~\ref{fig:5} and Eq.~\eqref{eq:kmlambda}, where $k_m$ are obtained from the transfer matrix, as a result of standard MPS truncation. }
 \label{fig:6}
\end{figure}

Qualitatively, this procedure boils down to dividing the continuous momenta $k\in(0,\pi)$ into windows which are logarithmically shrinking toward $k=0$ (minimum of $\epsilon^{\mathcal{T}_F}$), and representing each window by a single effective mode, as pictorially  presented in Fig.~\ref{fig:6}. The mode corresponding to the largest momentum represents the action of a few sites localized close to the impurity in the virtual direction (small $|l|$ in Fig.~\ref{fig:network}), while smaller momenta modes describe the relevant degrees of freedom which cannot be sharply localized around the impurity and are supported on sites extending to larger values of $|l|$. For gapped systems, this procedure can be terminated at the infrared cut-off related to the correlation length and a good approximation is obtained with a finite number of modes, resulting in a finite bond dimension.

Therefore, Fig.~\ref{fig:5} and Eq.~\eqref{eq:kmlambda} provide a direct evidence that such an interpretation of the origin of the transfer matrix obtained in the MPS algorithm is indeed correct.
This picture is further validated in Ref.~\onlinecite{Bal2015}, where a general tensor-network ansatz based on this idea is constructed.

The values of $\lambda$ in Eq.~\eqref{eq:kmlambda} depend both on the bond dimension $\chi$ and the correlation length in the system, i.e. $\xi \approx \Delta^{-1}$ in the gapped system and $\xi \approx L$ in the critical one. Qualitatively, one could expect that $k_1 \approx \frac{\pi}{\xi}$ and $k_{2\chi} \approx \pi$. 
We get rid of the proportionality constant appearing in Eq.~\eqref{eq:kmlambda} by considering the ratio $k_{2\chi}/k_{1} = \lambda^{2\chi-1} \approx \xi$ and thus expect that
\begin{equation}
\log \lambda \approx \frac{\log{\xi}}{2\chi}.
\label{eq:lambdachi}
\end{equation}
Quantitatively, we check this by extracting $\log \lambda$ (the slope of the linear dependence in Fig.~\ref{fig:5}) for various values of $\xi$ and $\chi$. While we see that Eq.~\eqref{eq:lambdachi} requires some corrections, it is capturing the {\it leading} tendency quite well. One of the reasons behind the correction could be, for example, that the smallest values of $k_{1,\dots}$ are visibly (in Fig.~\ref{fig:5}) affected by the infrared cut-off  and Eq.~\eqref{eq:kmlambda} does not describe them well. 


\subsection{Form factors resulting from the truncation}
\begin{figure}[t]
\begin{center}
\includegraphics[width=0.8\columnwidth]{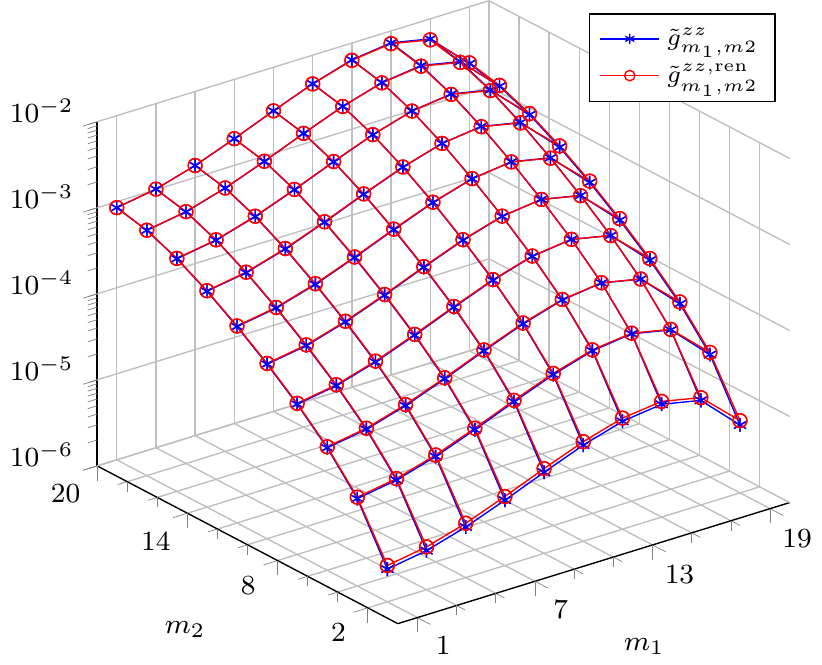}
\end{center}
\caption{(color online). Blue asterisks show the form factors $\tilde g^{zz}_{m_1,m_2}$ for the ZZ correlation function for the system with $g = 1.01$, $\gamma=0.8$ in the truncated state with $\chi=10$. Only unique, non-zero form factors are plotted.  For comparison, red circles show the results of a simple renormalisation procedure, where  $\tilde g^{zz,\mathrm{ren}}_{m_1,m_2}$ (see Eq.~\eqref{eq:rengzz}) is obtained by integration of the form factors before truncation over the momenta window surrounding $k_{m_1}$ and $k_{m_2}$, depicted pictorially in Fig.~\ref{fig:6}.
}
 \label{fig:7}
\end{figure}

We obtain further evidence in support of the interpretation presented in the previous section by looking at the form factors calculated for the truncated state. We focus on the ZZ correlation function and define the form factors
\begin{equation}
\tilde g^{zz}_{m_1,m_2} = ( \tilde \varphi_\emptyset \vert  \tilde{ \mathcal{T}}^{\sigma^z}  \vert \tilde  \varphi_{m_1,m_2} ) (\tilde  \varphi_{m_1,m_2}  \vert  \tilde{ \mathcal{T}}^{\sigma^z}  \vert \tilde  \varphi_\emptyset),
\label{eq:truncatedff}
\end{equation}
where $\tilde {\mathcal{T}}^{\sigma^z}= \bar{\tilde A}^{1}  \otimes \tilde A^1 - \bar{\tilde A}^{2}  \otimes \tilde A^2$ is the operator transfer matrix for $\sigma^z$,
$ \vert \tilde  \varphi_\emptyset)$ is the dominant eigenstate of the truncated transfer matrix $\tilde {\mathcal{T}}$ and $\vert \tilde  \varphi_{m_1,m_2}) = d_{m_1}^\dagger d_{m_2}^\dagger  \vert \tilde  \varphi_\emptyset)$ are eigenstates corresponding to two excited quasiparticles.
The form factors corresponding to single particle excitations are zero due to fermionic parity symmetry. We plot the unique, non-zero form factors $\tilde g^{zz}_{m_1,m_2}$ for a specific point with $g=1.01$ and $\gamma = 0.8$ in the paramagnetic phase, and  bond dimension $\chi=10$ in Fig. \ref{fig:7}. 
We point out, that the form factors most relevant for the long range correlations (i.e. corresponding to smallest single particle energies) have the smallest values in that plot. 

Additionally, we observe that the form factors are zero when the indices $m_1,m_2 = 1,2 \ldots 2 \chi$ are simultaneously odd or even, which can be attributed to the symmetry when reflecting bra and ket parts of the transfer matrix and we do not show them in the plot.  Both the details of the calculations and the above symmetry -- in the exact case where it is also present -- are discussed in the Appendix.

We argue that the values of the form factors can be understood as emerging through the renormalization group procedure described in Sec. IIIC. According to this each mode is effectively representing the window of momenta of the original -- not truncated -- transfer matrix $\mathcal{T}^F$, as depicted in Fig.~\ref{fig:6}. At the same time, the dominant, two-particle contribution to the correlation function is truncated as
$\frac12 \iint_{0}^{\pi} f^{zz}_{k_1,k_2} e^{-(\epsilon^{\mathcal{T}_F}_{k_1} + \epsilon^{\mathcal{T}_F}_{k_2}) R} \to \frac12 \sum_{m_1,m_2=1}^{2\chi} \tilde g^{zz}_{m_1,m_2} e^{-(\tilde \epsilon_{m_1} + \tilde \epsilon_{m_2}) (R-1)}$.
Consequently, we could expect that the form factor $\tilde g^{zz}_{m_1,m_2}$ is obtained as a sum of all the form factors $f^{zz}_{k_1,k_2} e^{-\epsilon^{\mathcal{T}_F}_{k_1}-\epsilon^{\mathcal{T}_F}_{k_2}}$ in that window (notice that $f^{zz}_{k_1,k_2}$ was defined using only  $\mathcal{Q}_z$, in the expression for the operator transfer matrix $\mathcal{T}_F^{\sigma^z} = \mathcal{Q}_z \mathcal{T}_F$, resulting in the additional factor $e^{-\epsilon^{\mathcal{T}_F}_{k_1}-\epsilon^{\mathcal{T}_F}_{k_2}}$ appearing above, and also that $f^{zz}_{k_1,k_2} \sim dk_1 dk_2$). 
In order to test this hypothesis we introduce,
\begin{equation}
\tilde g^{zz,\mathrm{ren}}_{m_1,m_2} = \int_{k_1 \in \Delta_{m_1}}  \int_{k_2 \in \Delta_{m_2}} f^{zz}_{k_1,k_2} e^{-\epsilon^{\mathcal{T}_F}_{k_1}-\epsilon^{\mathcal{T}_F}_{k_2}},
\label{eq:rengzz}
\end{equation}
where $\Delta_m$ represents the window around the momentum $k_m$ in Fig. \ref{fig:6}. As the most crude approximation, this means that the form factor would be proportional to the (logarithmically shrinking toward $k=0$) size of the corresponding momentum window. 

 The results of the above procedure are shown in Fig.~\ref{fig:7} reproducing the actual form factors with the relative error of maximally a couple of percent, which could be probably brought down even more by picking the exact size of the windows $\Delta_m$ in a more sophisticated way -- here we used $\Delta_m = (k_{m-1},k_{m+1})$ taking into account that roughly half of the form factors are zero. Still, even for such a simple procedure the agreement is exceptionally good as the form factors in Fig.~\ref{fig:7} span almost 4 orders of magnitude.

{\section{Conclusions}}
In this article, we constructed an exact MPS representation for the ground state of the XY model and showed how the Ornstein-Zernike form of the correlation function is naturally emerging in this picture. Subsequently we truncated this state (which has an exponentially-diverging MPS bond dimension), obtaining its approximation with relatively small bond dimension -- a procedure which is commonly employed in the representation of quantum states using tensor networks.

By analyzing this truncation, we can conclude, that an MPS  with finite bond dimension can be understood as a particular renormalization group procedure applied to the exact transfer matrix, whose dominant part is increasingly well reproduced with increasing bond dimension. The free-fermionic nature of the problems allows for a
 concise description of the state, and, for instance, its transfer matrix. The obtained spectrum of the MPS transfer matrix behaves as would be expected from the RG scheme based on a description in terms of an effective impurity -- proposed in Ref.~\cite{Zauner2014}. While the current discussion was limited to exactly solvable system, it motivates further studies of the intrinsic structure of the MPS matrices. Such an analysis is indeed done in \cite{Bal2015}, where a general tensor network algorithm based on the impurity picture, discussed in Sec. III.C, is designed. 

Likewise, the mapping employed in this article in Eq. \eqref{eq:XYmap} is limited to the commensurate phase of the XY model, where the transfer matrix is hermitian and most of our procedures were based on that fact. It would be interesting to extend the mapping into the incommensurate phase and non-hermitian transfer matrices, especially in context of testing Ref.~\cite{Zauner2014} which connects such oscillations with the position of minima of the dispersion relation of the Hamiltonian.


\section{Acknowledgements}
Discussions with Vid Stojevic, Viktor Eisler are gratefully acknowledged.  
We acknowledge support by NCN grant 2013/09/B/ST3/01603 (M.M.R.), EU grants SIQS and QUERG and the Austrian FWF SFB grants FoQuS and ViCoM (V.Z. and F.V.), and Research Foundation Flanders (J.H.).



\section*{Appendix}           
\renewcommand{\theequation}{A\arabic{equation}}
\setcounter{equation}{0}    
\setcounter{figure}{0}                                           
\renewcommand{\thefigure}{A\arabic{figure}}     

\subsection{Form factors and the correlation functions}
In this section we extend the discussion from Sec.~IIB by considering other correlation functions.  We numerically obtain the form factors, extract their relevant scaling and show how the asymptotic scaling of the correlation functions is emerging as a result. We compare these observations with the analytical results obtained in Ref. \cite{Barouch1971} by direct calculation of the correlation functions in the XY model -- an approach which is computationally significantly less complicated. Consequently, our analysis is intended just as an illustration of the underlaying mechanism.

As discussed is Sec. IIB, in our framework, the connected correlation function is calculated as:
\begin{equation}
C_{\hat o \hat o}(R)= \langle \hat o_0 \hat o_R \rangle - \langle \hat o_0 \rangle\langle \hat o_R \rangle =  \sum_{\alpha \neq \emptyset} f_\alpha^{oo} e^{ - E_\alpha R},
\end{equation}
where we consider local operators $\hat o = \sigma^x,\sigma^y,\sigma^z$, and the index $\alpha$ enumerates all the eigenstates of the Hermitian transfer matrix $\mathcal{T}_F$.
The form factors are defined as
\begin{equation}
f_\alpha^{oo} = \frac{1}{\mathcal{N}}\sum_{i=1}^{\mathcal{N}}( \varphi_\emptyset^i \vert \mathcal{Q}_o  \vert \varphi_\alpha ) ( \varphi_\alpha \vert \mathcal{Q}_o  \vert \varphi_\emptyset ^i),
\end{equation}
where $\mathcal{N}$ is the number of degenerated and orthogonal, dominant eigenvectors of $\mathcal{T}_F$, which we label as $|\varphi_\emptyset^i )$. The relevant, localized part of the operator transfer matrix, which we decompose as $\mathcal{T}_F^{\hat o} = \mathcal{Q}_o \mathcal{T}_F$, is
\begin{eqnarray}
\mathcal{Q}_z &=&  \exp \Br{-2 K_1 \tau_{-1} \tau_{1}}, \nonumber \\
\mathcal{Q}_x &=& \sqrt{1-e^{-4 K_1}} (\tau_{-1} +\tau_{1} )/2, \label {eq:Qxyz} \\
\mathcal{Q}_y &=& i \sqrt{e^{4 K_1}-1} (\tau_{-1} -\tau_{1} )/2, \nonumber
\end{eqnarray}
where $\tau_l = \cosh \overline K_2 \tau^x_l + i \sinh \overline K_2 \tau^y_l $ and following the convention in the main text where the position in the virtual direction $l$ is consistent with Fig.~\ref{fig:network}.

The operators $\mathcal{Q}_x$ and $\mathcal{Q}_y$, when mapped onto a fermionic system  as in the Appendix B, contain a string operator which extends over half of the chain. Therefore, we limit ourselves here to numerical calculations of the expectation values for some large but finite value of $L$.  The transfer matrix $\mathcal{T}_F$ is diagonalized as discussed in the next section, where its diagonal form is given by Eq.~\eqref{eq:diagonalTF} with the single particle energies given by Eq.~\eqref{eq:dispersionTF}, (almost) uniformly distributed  momenta  $k \in (0,\pi)$ (for non-zero $\epsilon^{\mathcal{T}_F}_k$) and the effective $dk = \pi/2 L$. 

The transfer matrix is invariant with respect to the transformation $l \to -l$. Likewise, $\mathcal{Q}_x$ and  $\mathcal{Q}_z$ are even with respect to that transformation and $\mathcal{Q}_y$ is odd. This suggest that the form factors related to the excited states with non-matching symmetry would be zero, which we indeed observe. For that reason one also expects that the mixed form factor $f^{xy}=0$. Similarly, $f^{xz} = f^{yz} = 0$ as $\mathcal{Q}_z$ is conserving fermionic parity and  $\mathcal{Q}_x$ and  $\mathcal{Q}_y$ are changing it. This is consistent with the fact that in the ground state of XY model $C_{xy}(R) = C_{yz}(R) = C_{zx}(R)  = 0$. 

Below, we summarize our observations for the remaining correlation functions in different phases, neglecting the fraction of the form factors which are equal zero.

\begin{figure}[t]
\begin{center}
\includegraphics[width=0.8\columnwidth]{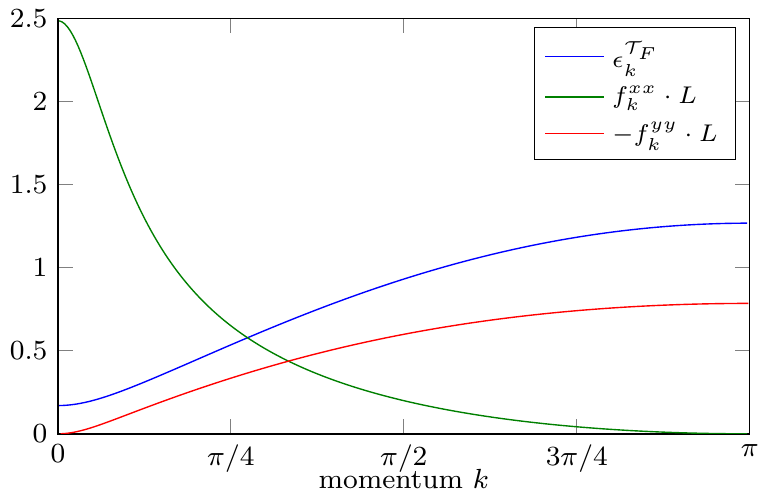}
\end{center}
\caption{(color online). Form factors for single-particle states, $f^{xx}_k$ and $f^{yy}_k$, and the single particle energies $\epsilon^{\mathcal{T}_F}_k$, for $g=1.1$ $\gamma=0.5$ in the paramagnetic phase. }
 \label{fig:ff_paramagnetic}
\end{figure}

{\it Paramagnetic phase.---}  We simulate the system for $(g,\gamma)=(1.1,0.5)$.
From the numerics, we observe that the non-zero form factors, which correspond to single-particle band, behave as $f^{xx}_k \simeq {\rm const} \cdot dk $ and $f^{yy}_k \sim k^2 ~ dk$ around $k=0$, and the dispersion relation $\epsilon^{\mathcal{T}_F}_k \simeq \Delta + a_p k^2$. We show the form factors in Fig.~\ref{fig:ff_paramagnetic}. The asymptotic form of the correlation functions is then determined by the single-particle band, as
\begin{eqnarray*}
C_{xx}(R) \sim \int_0^{\infty} dk \ e^{-(\Delta + a_p k^2) R} \sim \frac1{R^{1/2}} e^{-R \Delta}, \\
C_{yy}(R) \sim \int_0^{\infty} dk \ k^2 e^{-(\Delta+a_p k^2) R} \sim \frac{1}{R^{3/2}} e^{-R \Delta},  
\end{eqnarray*}
in agreement with \cite{Barouch1971}.

{\it Ferromagnetic phase.---} We simulate the system for $(g,\gamma)=(0.8,0.8)$. There is a single mode with $\epsilon^{\mathcal{T}_F}_\emptyset =0$  and
the dominant eigenstate of the transfer matrix is degenerated with $\mathcal{N}=2$. The rest of the spectrum is still given by Eq.~\eqref{eq:dispersionTF} with $\epsilon^{\mathcal{T}_F}_k \simeq \Delta + a_f k^2$ around the minimum at $k=0$. 
We observe that the form factors corresponding to single-particle band are zero, $f^{xx}_k = f^{yy}_k=0$, and likewise $f^{zz}_{k,\emptyset}=0$. 

For XX and YY correlations, the first nonzero form factors come from 3-quasiparticles excitations including $\epsilon^{\mathcal{T}_F}_\emptyset =0$, i.e. $\alpha = \{k_1,k_2,\emptyset\}$. From the specific example the data are consistent with the quadratic scaling of $f^{xx}_{k_1,k_2} \sim (k_1^2+k_2^2) dk_1 dk_2$ and quartic 
scaling of $f^{yy}_{k_1,k_2} \sim (k_1-k_2)^4 dk_1 dk_2$. Likewise for ZZ correlation, we see quadratic scaling $f^{zz}_{k_1,k_2} \sim (k_1^2+k_2^2) dk_1 dk_2$.
By performing double integrals like in the main text, this would lead to $C_{xx}(R) \sim \frac1{R^{2}} e^{-R 2 \Delta}$, $C_{yy}(R)  \sim \frac{1}{R^{3}} e^{-R 2 \Delta}$ and 
$C_{zz}(R) \sim  \frac1{R^{2}} e^{-R 2 \Delta}$ in agreement with \cite{Barouch1971}.

Notice that none of the correlation functions is falling off exponentially with the correlation length suggested by the transfer matrix $\xi = 1/\Delta$, but actually twice as fast. Interestingly, it is possible to recover the actual correlation length by adding the string operator between the two end points of the XX or YY correlation function, namely by considering $\langle \sigma^{x(y)}_0 \prod_{n=1}^{R-1}\sigma^{z}_n\sigma^{x(y)}_R\rangle$.

{\it Critical point for g=1.---} We simulate the system for $(g,\gamma)=(1,0.5)$. The ZZ correlation function, discussed in the main text, follows the scheme presented above, as the specific form of the $\mathcal{Q}_z$ makes all form factors, expect for the ones corresponding to the two-quasiparticle excitations, equal zero.
For the XX and YY correlations this is no longer the case, as the gap is equal zero ($\epsilon^{\mathcal{T}_F}_k \simeq a_c k$) and the contributions from many-quasiparticles-bands might be relevant for the algebraic scaling. We see that this is indeed the case. The data for growing $L$ suggest the scaling $f^{xx}_k \sim \frac{1}{k L^{0.25}} dk$, and $f^{yy}_k \sim - \frac{k}{L^{0.25}} dk$. Due to additional factors of $L^{-0.25}$ the contribution from the single-particle band is vanishing in the limit of $L \to \infty$. In order to recover the actual asymptotic behavior $C_{xx}(R) \sim  \frac1{R^{1/4}}$ and $C_{yy}(R) \sim  \frac1{R^{9/4}}$ \cite{Barouch1971}, the contribution coming from all multi-particle bands would have to be taken into account and the simple single-particle picture presented in this section does no longer apply.

\
\renewcommand{\theequation}{B\arabic{equation}}
\setcounter{equation}{0}    
\setcounter{figure}{0}                                           
\renewcommand{\thefigure}{B\arabic{figure}}     

\subsection{Free-fermionic description of transfer matrices}

{\it Transfer matrix.--- } We define the transfer matrix as $\mathcal{T}_F  = \sum_{i=0}^1 \bar{A}^{i} \otimes A^{i}$, where, for the XY model, matrices $A^i$ are given by Eq.~\eqref{eq:A}, resulting in
\begin{eqnarray}  
      &\mathcal{T}_F = W_1^{\frac12} W_2 W_1^{\frac12}, \\
      &W_1 = \exp\Br{\overline K_2 \sum_{l=1}^{2L}  \tau^z_l};  \ W_2 = \exp\Br{ K_1 \sum_{l=1}^{2L-1} \tau^x_l \tau^x_{l+1}}. \nonumber
\end{eqnarray}
We have reindexed the auxiliary spins along the vertical direction for convenience, so that sites with $l=1,2,\dotsc L$ correspond to $A^i$ and sites with $l=L+1,\dotsc, 2 L$ to $\bar{A}^{i}$. 
Operators of this form were diagonalized by Abraham \cite{Abraham1971} using the formalism of transformation matrices. Here we reiterate the main steps of the derivation.

Firstly, $\mathcal{T}_F$ is mapped onto a free-fermionic model by means of a Jordan-Wigner transformation $\tau_n^z = 1-2 c_n^\dagger c_n$, $\tau^x_n + i \tau^y_n = 2 c_n \prod_{m<n} \left( 1-2 c^\dagger_m c_m \right)$, where $c_n$ are fermionic annihilation operators. 
It is convenient to introduce Majorana fermions $c^M_{2n-1} = c_n + c_n^\dagger$ and $c^M_{2n} = i( c_n  - c_n^\dagger)$, with $\{c^M_m,c^M_n\}=2 \delta_{m,n}$, where we will use superscript $^M$ to indicate Majorana fermions.
Now, 
\begin{eqnarray*}  
   &W_1 = \exp\Br{\overline K_2 \sum_{l=1}^{2L}  i c^M_{2l-1} c^M_{2l}},  \\
   &W_2 = \exp\Br{ K_1 \sum_{l=1}^{2L-1} i c^M_{2l} c^M_{2l+1}}. 
\end{eqnarray*}

\noindent We define a (row) vector $\vec c^M = (c^M_1,c^M_2,c^M_3, \ldots, c^M_{4L})$ and for an operator $T$, which is an exponential of a free-fermionic Hamiltonian, we consider the similarity transformation
\begin{equation}
\label{eq_a_similarity}
T \vec c^M T^{-1} = \vec c^M R[T],
\end{equation}
which defines a $4 L \times 4 L $ transformation matrix $R[T]$. Above, it is understood that $T \vec c^M T^{-1} = (T c^M_1 T^{-1}, T c^M_2 T^{-1} , \ldots, T c^M_{4L} T^{-1})$.

It is convenient to introduce a $2\times2$ matrix
\begin{equation*}
u(x)= 
\begin{pmatrix}
\cosh  x & i \sinh  x \\
- i \sinh  x & \cosh  x \\
\end{pmatrix}.
\end{equation*}
The transformation matrices for $W_1^{\frac12}$ and $W_2$ are block diagonal 
\begin{eqnarray*}
 & R[W_1^{\frac12}] = \bigoplus_{n=1}^{2L} u(\overline K_2), \\
 & R[W_2] = 1 \oplus \left( \bigoplus_{n=1}^{2L-1} u(2 \overline K_1) \right) \oplus 1.
 \end{eqnarray*}
The transformation matrix for $\mathcal{T}_F$ is found simply by multiplying transformation matrices for $W_1^{\frac12}$ and $W_2$
\begin{equation}
\label{eq:RTF}
R[\mathcal{T}_F] = R[W_1^{\frac12}] R[W_2] R[W_1^{\frac12}].
\end{equation}
Subsequently, the transfer matrix is diagonalized by finding $U_\mathcal{T} \in SO(4L)$ which brings $R[\mathcal{T}_F]$ into canonical form 
\begin{eqnarray*}
& R[\mathcal{T}_F] = U_\mathcal{T} R_a^M U_\mathcal{T}^T, \\
& R_a = \bigoplus_{n=1}^{2L} u(\epsilon^{\mathcal{T}_F}_n).
\end{eqnarray*}
This gives single particle energies $\epsilon^{\mathcal{T}_F}_n>0$ and  $U_\mathcal{T}$ defines a new fermionic basis $\vec a^M = \vec c^M U_\mathcal{T}$, for which (up to normalization)
\begin{equation*}
 \mathcal{T}_F = \exp \Br{i \sum_{n=1}^{2L} \frac12 \epsilon_n^{\mathcal{T}_F} a^M_{2n-1} a^M_{2n}} = \exp \Br{-\sum_{n=1}^{2L}\epsilon_n^{\mathcal{T}_F} a_n^\dagger a_n}.
\end{equation*}
 The dominant eigenvector of $\mathcal{T}_F$ is annihilated by all annihilation operators $a_n| \psi_\emptyset ) =(a^M_{2n-1} - i a_{2n}^M) | \psi_\emptyset ) =0$.

{\it Reduced density matrix.--- } The reduced density matrix $\rho$ of $|\psi_\emptyset )$ with support on sites $l=1,2,\dotsc,L$ is diagonalized following standard techniques \cite{Peschel2009} by considering a $2L\times 2L$ covariance matrix for a half-chain 
\begin{equation*}
C_{[1,2,\dotsc,L]} = \Br{(\psi_\emptyset| c^M_m c^M_n  | \psi_\emptyset)}_{m,n = 1,2, \dotsc,2L} = \mathbb{1} + i B^{M},
\end{equation*}
where $B^{M}$ is skew-symmetric. The reduced density matrix is diagonalized by bringing $B^M$ into canonical form
\begin{eqnarray*}
& B^M = U_B B_f^M U_B^T, \\
& B^M_f = \bigoplus_{n=1}^L \begin{pmatrix}
0 &  - \tanh \delta_n \\
 \tanh \delta_n  & 0 \\
\end{pmatrix},
\end{eqnarray*}
where $\delta_m$ defines the entanglement spectrum. $U_B \in SO(2L)$ defines a new set of Majorana fermions $(f_1^M,\dotsc,f_{2L}^M) =  (c_1^M,\dotsc,c_{2L}^M) U_\mathcal{B}$, for which 
\begin{equation*}
 \rho = \frac{1}{Z'} \exp \Br{i \sum_{n=1}^{L} \delta_n f^M_{2n-1} f^M_{2n}} =\frac{1}{Z} \exp \Br{- 2 \sum_{n=1}^{L} \delta_n f^\dagger_n f_n},
\end{equation*}
where $Z$ and $Z'$ are the normalization factors.
Similarly, we obtain $(f^M_{2L+1},f^M_{2L+2},\dotsc,f^M_{4L})$ by considering the reduced density matrix supported on sites $L+1,\dotsc,2L$.

{\it Truncation.---}  Finally, we describe how to obtain the truncated transfer matrix $\tilde {\mathcal{T}}=\sum_{i=1}^2  \bar{\tilde A}^{i}  \otimes \tilde A^i$, where 
$\tilde A^i$ are given by Eq.~\eqref{eq:Anew} and we keep the $\chi$ most relevant fermionic modes, i.e. we discard $f$-modes for  $\Lambda = \{\chi+1,\chi+2,\dotsc,L\} \cup \{ L+\chi+1,L+\chi+2,\dotsc ,2 L\}$. 

We work directly with the transfer matrix and find 
$$
\tilde{\mathcal{T}} = (\otimes _{j \in \Lambda } (0_j|)  \mathcal{T}^F  (\otimes_{j \in \Lambda }| 0_j)),$$ 
where states $|0_j)$, for which $f_j|0_j) =0$, are obtained from the diagonalization of the reduced density matrix. We use the formalism of the transformation matrix, extending it to the case of non-invertible projections, cf. Eq.~\eqref{eq_a_similarity}.

First, we obtain the transformation matrix for $\mathcal{T}_F$ in the $f$-fermionic base (it is convenient to work with Dirac fermions here) as
\begin{equation*}
\mathcal{T}^F \vec f \mathcal{T}^F = \vec f R[\mathcal{T}^F]^f,
\end{equation*} 
where for convenience we reorder  
$\vec f =\{\vec f_a, \vec f_b, \vec f_c \} $ with $\vec f_a$ describing the relevant modes $\vec f_a = \{f_j, f_j^\dagger : j \notin \Lambda \}$, $\vec f_b = \{f_j : j \in \Lambda\}$  are annihilation operators corresponding to the truncated modes, and finally  $\vec f_c = \{f_j^\dagger : j \in \Lambda\}$ denote the corresponding creation operators. $R[\mathcal{T}^F]^f$ can be directly obtained from $R[\mathcal{T}^F]$ in Eq.~\eqref{eq:RTF} by a suitable basis rotation from $\vec a^M$ to $\vec f$. 

\noindent Here, the relevant sub-matrices of $R[\mathcal{T}^F]^f$ are 
\begin {align*}
R_{aa} &= \Br{R[\mathcal{T}^F]^f}_{m,n = 1,\dotsc 4 \chi} \\
R_{ab} &= \Br{R[\mathcal{T}^F]^f}_{m = 1,\dotsc 4 \chi, n = 4\chi +1, \dotsc, 2 L + 2 \chi } \\
R_{ba} &= \Br{R[\mathcal{T}^F]^f}_{m = 4\chi +1, \dotsc 2 L + 2 \chi, m = 1,\dotsc 4 \chi} \\
R_{bb} &= \Br{R[\mathcal{T}^F]^f}_{m,n = 4\chi +1, \dotsc 2 L+2 \chi} 
\end {align*}
Namely, $R_{aa}$ describes transformation of  $\vec f_a$  into  $\vec f_a$ under the similarity transformation given by $\mathcal{T}^F$,
 $R_{ab}$  corresponds to the transformation $\vec f_a$  into  $\vec f_b$, etc.
 
The transformation matrix corresponding to $\tilde {\mathcal{T}}$, 
\begin{equation}
\label{eq:tT}  
\tilde{ \mathcal{T}} \vec f_a \tilde{ \mathcal{T}}^{-1} = \vec  f_a R [\tilde{\mathcal {T}} ],
\end{equation}
is found as
\begin{equation}
\label{eqa:Rtruncate}  
R[\tilde{\mathcal{T}}] = R_{aa} - R_{ab} R_{bb}^{-1} R_{ba}. 
\end{equation}
Now, bringing $R[\tilde{\mathcal{T}}]$ into canonical form -- similar to $R[{\mathcal{T}^F}]$ -- yields the spectrum $\tilde \epsilon_m$, where
$$
\tilde {\mathcal{T}} = \exp \Br{\sum _{m=1}^{2\chi} \tilde \epsilon_m d_m^\dagger d_m }. 
$$

In order to derive equation Eq.  \eqref{eqa:Rtruncate} we consider 
\begin{equation}
\label{eq:PTP}
\hat P \mathcal{T}^F \vec f \hat P = \hat P \vec f \mathcal{T}^F \hat P R[\mathcal{T}^F]^f,
\end{equation}

\noindent where the projection $P_\Lambda =\prod_{j\in\Lambda}|0_j)(0_j|= \prod_{j\in\Lambda} c_j c_j^\dagger$. 
Notice that $\vec f_b \hat P =0$ and $\hat P \vec f_c  =0$.  
Rewriting Eq.~\eqref{eq:PTP} we obtain 
\begin{eqnarray*}
\hat P \mathcal{T}^F \hat P \vec f_a &=& \vec f_a \hat P \mathcal{T}^F \hat P R_{aa} + \hat P \vec f_b\mathcal{T}^F \hat P R_{ba}, \\
0 &=& \vec f_a \hat P \mathcal{T}^F \hat P R_{ab} +\hat P \vec f_b\mathcal{T}^F \hat P R_{bb}. 
\end{eqnarray*}
Eliminating $\hat P \vec f_b\mathcal{T}^F \hat P$ from the above equation leads to 
$$
\hat P \mathcal{T}^F \hat P \vec f_a  =  \vec f_a \hat P \mathcal{T}^F \hat P (R_{aa}  - R_{ab} R_{bb}^{-1} R_{ba}).
$$
Now it is enough to notice that $\hat P \mathcal{T}^F \hat P \sim  \hat  P \tilde{ \mathcal{T}} \hat P$ and since $\hat P$ works nontrivially only on modes $f_j$ with $j \in \Lambda$, and  $\tilde {\mathcal{T}}$ on modes with $j \notin \Lambda$ we obtain Eq.~\eqref{eq:tT} with $R[\tilde {\mathcal{T}}]$ given by Eq.~\eqref{eqa:Rtruncate}.

{\it Form factors in the truncated state.---} We focus on the form factors for the ZZ correlation function. The (full) operator transfer matrix is
\begin{equation}  
      \mathcal{T}_F^{\sigma^z} = W_1^{\frac12} \exp \left[-2 K_1 \tau^x_L \tau^x_{L+1}\right] W_2 W_1^{\frac12}, 
\end{equation}
and after mapping onto free-fermionic system we describe it using the transformation matrix $R[\mathcal{T}_F^{\sigma^z} ]$ in an analogous way to the transfer matrix $\mathcal{T}_F$.  We work directly with this transfer matrix and calculate its form after truncation by using Eqs. \eqref{eq:tT} and \eqref{eqa:Rtruncate} finding $R[\tilde{\mathcal{T}}^{\sigma^z}]$ in the process.
Now, in order to find the form factors in Eq.~\eqref{eq:truncatedff} we rewrite
\begin{equation}
(\tilde \varphi_\emptyset \vert  \tilde{ \mathcal{T}}^{\sigma^z}  \vert \tilde  \varphi_{m_1,m_2} ) = \Tr\left[\tilde{ \mathcal{T}}^{\sigma^z} d_{m_1}^\dagger  d_{m_2}^\dagger d_1 d_1^\dagger \ldots    d_L d_L^\dagger  \right],
\label{eq:fffromfermions}
\end{equation}
where $d_1 d_1^\dagger \ldots    d_L d_L^\dagger$ is projector onto the dominant eigenstate of $\tilde{\mathcal{T}}$.  We can think about $\tilde{ \mathcal{T}}^{\sigma^z}/\Tr(\tilde{ \mathcal{T}}^{\sigma^z})$ as a reduced density operator and using Wick's theorem calculate the above expression (up to normalization) as a Pfaffian of the two-point correlation matrix $\Tr\left(\tilde{ \mathcal{T}}^{\sigma^z} b_1 b_2 \right)$, where $b_1,b_2 = d_1,d_1^\dagger \ldots d_L, d_L^\dagger$ are all  pairs of $d$ operators appearing in Eq.~\eqref{eq:fffromfermions}. They, in turn, can be easily calculated by finding the canonical form of $R[\tilde{\mathcal{T}}^{\sigma^z}]$, calculating the two point correlations in that base and subsequently rotating them into $d$ fermions.

Using this approach, we have to reintroduce the normalization $\Tr(\tilde{ \mathcal{T}}^{\sigma^z})$ by hand. Here, we do it by calculating the magnetization  
$M_z = (\tilde \varphi_\emptyset \vert  \tilde{ \mathcal{T}}^{\sigma^z}  \vert \tilde  \varphi_\emptyset )$, as discussed above, and  comparing it with the exact value.


\begin{thebibliography}{99}
\bibitem{normalization_comment} {For clarity we neglect the -- in this case irrelevant  -- normalization factors throughout the article and reintroduce them only when necessary}.
\bibitem{PBC_comment} {In the previous section we work with the system which has periodic boundary conditions, but, in the thermodynamic limit of $N\to \infty$, since the entanglement spectrum of large enough block is effectively doubled, we perform truncation at each site locally as in the case of system with open boundary conditions.
}
\bibitem{iDMRG*}  We obtain an MPS for some larger bond dimension $D'$ and subsequently truncate to $D=2^\chi$ by keeping only Schmidt values corresponding to Eq.~\eqref{eq:newrho}. They can be identified by finding groups of Schmidt values which have constant ratios (with reasonable precision), resulting from the structure of Eq.~\eqref{eq:newrho}.
\end{thebibliography}
\end{document}